\documentclass[twocolumn,floatfix,prl,aps,showpacs]{revtex4-1}
\usepackage{graphicx,color,amsmath,amssymb}

\newcommand{\be}{\begin{equation}}
\newcommand{\ee}{\end{equation}}
\newcommand{\ba}{\begin{eqnarray}}
\newcommand{\ea}{\end{eqnarray}}

\begin{document}

\title{Enigmatic 4/11 State: A Prototype for Unconventional Fractional Quantum Hall Effect}
\author{Sutirtha Mukherjee,$^1$ Sudhansu S. Mandal,$^1$  Ying-Hai Wu,$^2$ Arkadiusz W\'ojs,$^3$  
and Jainendra K. Jain$^{2}$}
\affiliation{$^{1}$Department of Theoretical Physics, Indian Association for the Cultivation of Science, 
         Jadavpur, Kolkata 700 032, India}
\affiliation{ $^{2}$Department of Physics, 104 Davey Lab, Pennsylvania State University, University Park PA, 16802}
\affiliation{
   $^{3}$Institute of Physics,  Wroclaw University of Technology, 50-370 Wroclaw, Poland}

\begin{abstract}
The origin of fractional quantum Hall effect (FQHE) at 4/11 and 5/13 has remained controversial. We make a compelling case that FQHE is possible here for fully spin polarized composite fermions, but with an unconventional underlying physics. Thanks to a rather unusual interaction between composite fermions, FQHE here results from the suppression of pairs with relative angular momentum {\em three} rather than one, confirming the exotic mechanism proposed by W\'ojs, Yi and Quinn [Phys. Rev. B {\bf 69}, 205322 (2004)]. We predict that the 4/11 state reported a decade ago by Pan {\em et al.} [Phys. Rev. Lett. {\bf 90}, 016801 (2003)] is a conventional partially spin polarized FQHE of composite fermions, and estimate the Zeeman energy where a phase transition into the unconventional fully spin polarized state will occur. 
\end{abstract}

\maketitle

The fractional quantum Hall effect (FQHE) \cite{Tsui82} is one of the cleanest and most nontrivial manifestations of interelectron interaction, and has produced a string of surprising discoveries during the last three decades. A FQHE state is characterized by $(f, \gamma, \alpha)$, where $f$ is the fraction appearing in the expression for the fractionally quantized Hall resistance $R_{\rm H}=h/f e^2$ (indicating an incompressible state at filling factor $\nu=f$), $\gamma$ is the spin/valley polarization, and $\alpha$ labels topologically distinct states with the same $f$ and $\gamma$ that may occur for different interactions. The richness of the physics is made evident by the remarkable fact that more than 75 fractions have been observed to date \cite{WeiPan}, and states with several different spin/valley polarizations occur at many of these fractions. Different physical mechanisms for FQHE have been identified. Many FQHE states at filling factors of the form $\nu=j\pm n/(2pn\pm 1)$, where $j$, $n$ and $p$ are integers, are explained as integer quantum Hall effect (IQHE) of composite fermions carrying $2p$ vortices \cite{Jain89},  and the FQHE states at $\nu=5/2$ and 7/2 are modeled as chiral p-wave paired states of composite fermions \cite{MooreRead1991}. Our focus here is on the FQHE at $\nu=4/11$ and 5/13 \cite{Pan03} which cannot be understood as either IQHE or paired state of composite fermions. We show below that their explanation requires yet another physical mechanism, thus adding to the richness of the FQHE and opening the exciting possibility of other FQHE states arising from this mechanism. 

The 4/11 and 5/13 FQHE states are very delicate, appearing only in the highest quality samples \cite{Pan03}; in fact, a definitive observation, in the form of accurately quantized Hall plateaus with activated longitudinal resistance, is still lacking.  These states were seen at fairly large magnetic fields ($\sim$ 11T) where the Zeeman splitting ($E_{\rm Z}$) is substantial, $\sim$ 3K, and the resistance showed negligible variation upon increase in $E_{\rm Z}$; these facts were taken in Ref.~\cite{Pan03} strongly to support a fully spin polarized FQHE. We will therefore look for a fully spin polarized state at these fractions, returning to the role of spin later. In this filling factor region, electrons capture two quantized vortices each to turn into composite fermions \cite{Jain89}. Composite fermions experience an effective magnetic field $B^*=B-2\phi_0\rho$, where $B$ is the external field, $\phi_0=hc/e$ is the flux quantum, and $\rho$ is the electron or composite fermion (CF) density. Composite fermions form Landau-like levels called $\Lambda$ levels ($\Lambda$Ls) in $B^*$, and their filling factor $\nu^*$ is related to the electron filling factor $\nu$ by the relation $\nu=\nu^*/(2\nu^*\pm 1)$. The IQHE of composite fermions at $\nu^*=n$ manifests as FQHE at odd-denominator fractions of the form $\nu=n/(2n\pm 1)$. These will be referred to as the ``conventional" FQHE states.

At 4/11 and 5/13 the CF filling is $\nu^*=1+1/3$ and $\nu^*=1+2/3$, and the question is what state composite fermions form at 1/3 and 2/3 filling in the second $\Lambda$L. Several proposals have been made, but all are subject to criticisms. A variational study \cite{LeeScarola} suggested that they form a crystal, while another \cite{Goerbig04} suggested a conventional Laughlin-type \cite{Laughlin1983} FQHE state. The wave functions employed in these studies, however, have not been demonstrated to be sufficiently accurate to capture the subtle physics of this state. Ref.~\cite{Chang04} performed CF diagonalization \cite{Mandal1,JainKamilla1} and also supported the conventional FQHE, primarily based on results for the 12 particle system; this system, however, was recently recognized \cite{MukherjeeMandalWojs2012} to ``alias" with the anti-Pfaffian paired state at $\nu=3/8$, thereby casting doubt on the conclusions of Ref.~\cite{Chang04}. W\'ojs, Yi and Quinn (WYQ) \cite{Wojs1,Wojs2} modeled composite fermions in the second $\Lambda$L as fermions interacting via an effective 2-body interaction, which is determined by placing two composite fermions in the second $\Lambda$L \cite{Sitko,LeeScarola,Lee1}. They studied the effective model by exact diagonalization and arrived at the surprising conclusion that  composite fermions form ``unconventional" 1/3 and 2/3 states. However, the 2-body model is known sometimes to produce a wrong ground state \cite{comment1}, presumably because of the neglect of either 3 and higher body interaction between composite fermions, or the filling factor dependence of the inter-CF interaction. The situation therefore remained unsettled.

Which state is energetically favored is determined by the very weak interaction between composite fermions. Fortunately, the method of CF diagonalization \cite{Mandal1} (CFD) has been shown to capture the physics of inter-CF interaction extremely accurately in the region of interest, producing energies within $\sim$ 0.05\% of the exact energies. In this method, a correlated CF basis \{$\Psi^{{\rm CF},\alpha}_{\nu}$\} is constructed starting from the known basis \{$\Phi^{\alpha}_{\nu^*}$\} of degenerate ground states of noninteracting fermions at $\nu^*$, and then the full Coulomb Hamiltonian is diagonalized within this basis. The basis functions $\Psi^{{\rm CF},\alpha}_{\nu}$ are much more complicated than the usual Slater determinants, but efficient methods have been developed to calculate with them \cite{JainKamilla1,Mandal1}. The dimension of \{$\Psi^{{\rm CF},\alpha}_{\nu}$\} is exponentially small compared to the dimension of the full lowest Landau level (LLL) Hilbert space, which allows CFD to treat much larger systems than possible for exact diagonalization. We stress that no assumption is made regarding the form of the interaction between composite fermions.  More details can be found in Supplemental Material (SM) \cite{SM}.

\begin{center}
\begin{table}[b]
\centering
\begin{tabular}{|c||c|c|}
\hline
$\nu$& conventional & unconventional\\ 
\hline
\hline
1/3 & $S^*=3$ & $S^*=7$ \\ \hline
2/3 & $S^*=0$ & $S^*=-2$ \\ \hline
4/11 & $S=4$ & $S=5$ \\ \hline
5/13 & $S=17/5$ & $S=13/5$ \\
\hline
\end{tabular}
\caption{Shifts for the conventional and unconventional states at 1/3, 2/3, 4/11 and 5/13.}
\label{Shifts}
\end{table}
\end{center}

We use the spherical geometry \cite{Haldane1}, in which $N$ electrons move on the surface of a sphere under the influence of a flux of $2Q(hc/e)$, where $2Q$ is an integer. The many particle eigenstates are labeled by the total orbital angular momentum $L$.  Theoretical demonstration of incompressibility at a filling $\nu$ requires that: the state at each $N$ and $2Q$ satisfying  $2Q=\nu^{-1}N-S$, where $S$ is an $N$ independent ``shift," produces a uniform ($L=0$) state separated from the excitations by a gap, and the gap extrapolates to a nonzero value in the limit $N\rightarrow \infty$. Candidate states with different values of $S$ at a given $\nu$ are topologically distinct, and a determination of $S$ by exact or CF diagonalization can identify which candidate state is viable. The shifts $S^*$ for the conventional and WYQ states at 1/3 and 2/3 are given in Table \ref{Shifts}; these result in 4/11 and 5/13 states at shifts $S$ shown in Table \ref{Shifts} (see SM for details \cite{SM}).

\begin{figure}[t]
\includegraphics[width=0.5\textwidth]{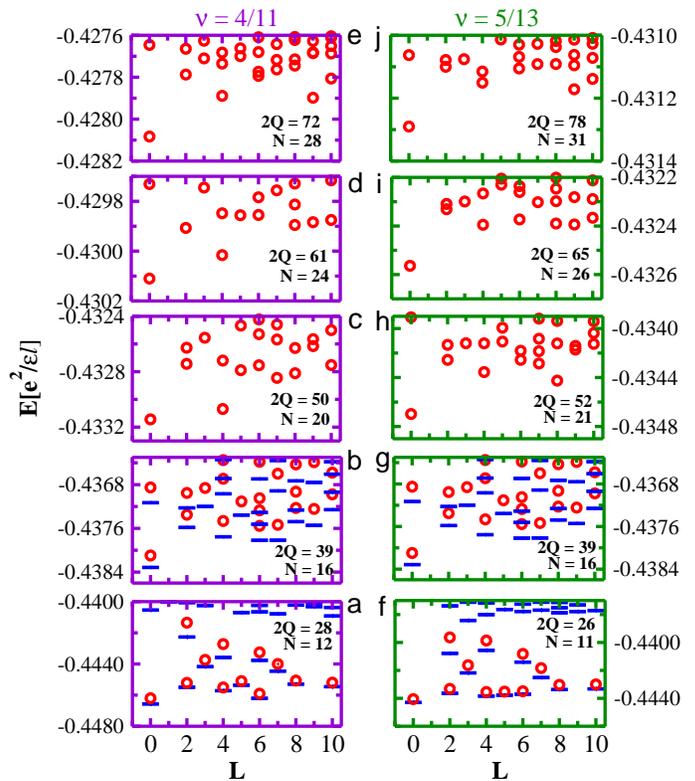}
\caption{Composite fermion spectra at 4/11 and 5/13. The circles show energies {\em per particle} obtained by CF diagonalization for the FQHE state at 4/11 (left) and 5/13 (right) at shifts $S=5$ and 13/5, respectively. The dashes are obtained by exact diagonalization of the Coulomb interaction in the full LLL Hilbert space (only the very low energy states are shown; the dimensions of the full LLL basis and the CFD basis are given in SM \cite{SM}). $N$ is the number of particles, $2Q$ is the number of flux quanta passing through the sample, and $L$ is the total angular momentum quantum number of the eigenstate. The energy per particle is quoted in units of $e^2/\epsilon \ell$, where $\ell=\sqrt{\hbar c/eB}$ is the magnetic length and $\epsilon$ is the dielectric constant of the host material. The energy includes the interaction with the positively charged neutralizing background. The $N=16$ state at $2Q=39$ occurs for both 4/11 and 5/13; we have included this panel twice for symmetry reasons.} \label{411}
\end{figure}

\begin{figure}[t]
\includegraphics[width=0.4\textwidth]{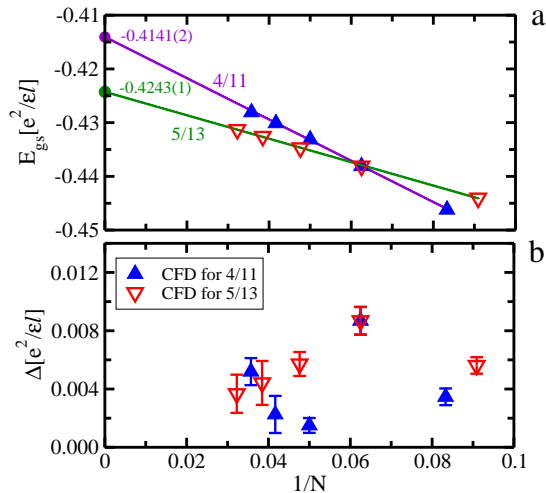}
\caption{(a) The ground state energy per particle for 4/11 and 5/13 as a function of $1/N$, extrapolated to the thermodynamic limit. (b) The energy gap of the lowest neutral excitation.} \label{gaps}
\end{figure}

The CFD spectra at the unconventional shifts are shown in Fig.~\ref{411}. Several points are noteworthy. (i) The comparison with exact spectra, available for up to 16 particles (Fig.~\ref{411}), demonstrates that the CFD spectra are to be treated as essentially exact for the ground states. (The CFD energies deviate from the exact ones by $\sim$ 0.05\%.) (ii) The ground state occurs at $L=0$ for each value of $N$ at the unconventional shifts. (iii) A reliable extrapolation of gaps to the thermodynamic limit is unfortunately not possible due to strong finite size effects, but the results are consistent with a nonzero value (see Fig.~\ref{gaps} for the energy of the lowest neutral excitation). The energy scale for the gaps (Fig. \ref{gaps}) is $\sim$0.002 $e^2/\epsilon \ell$, where $\ell=\sqrt{\hbar c/eB}$ is the magnetic length. This energy is approximately $\sim$50 times smaller than the ideal theoretical gap of 1/3 (0.1 $e^2/\epsilon \ell$), indicative of a much weaker interaction between composite fermions than that between electrons. (iv) The 2-body interaction model of WYQ would produce identical spectra for the horizontally neighboring panels in Fig.~\ref{411}. Substantial differences seen in the CFD spectra indicate the importance of the 3 and higher body interactions between composite fermions. (v) Finally, it is interesting to note that for the 4/11 state with 12, 16 and 28 particles,  the full dimensions of the $L=0$ sector (in the LLL) are 902, 250256 and $\sim 2\times 10^{13}$, respectively, whereas the dimensions of the corresponding CF bases are 1, 3 and 28.

We next show, by constructing explicit trial wave functions, that the origin of this FQHE is captured by the mechanism proposed by WYQ. A pairwise interaction for fermions confined to any Landau level (LL) can in general be parameterized as \cite{Haldane1} 
\be 
V=\sum_m V_m|m\rangle \langle m|
\label{m-3}
\ee
where $|m\rangle$ denotes the two particle state with relative angular momentum $m$, and the pseudopotential $V_m$ is the energy of this state. For the Coulomb interaction $V_1$ dominates, and the conventional states $n/(2n\pm 1)$ are produced in a model $V_m=\delta_{m,1}$. WYQ consider instead the model interaction $V_m=\delta_{m,3}$ and find, by numerical diagonalization, that it produces incompressible $L=0$ ground states at 1/3 and 2/3, but at shift 7 and -2, respectively, as opposed to the conventional shifts of 3 and 0 produced by the Coulomb (or $V_m=\delta_{m,1}$) interaction. The WYQ states are thus topologically distinct from Laughlin's 1/3 and 2/3 states. Explicit wave functions for them are not known, but can be generated exactly for up to 15 particles by a brute force numerical diagonalization. The WYQ states do not have zero energy, implying that they minimize, but do not eliminate, occupation of pairs with relative angular momentum 3. Given that there is no repulsion in the angular momentum $m=1$ channel, one might expect pairing correlations, but the actual FQHE state does not involve pairing, as evidenced by the fact that an incompressible state is produced for both even and odd $N$.
We have performed an extensive investigation of the 1/3 WYQ state through exact diagonalization on sphere, torus and disk, as well as through its entanglement spectrum. These studies, reported in the SM \cite{SM}, clarify that:  its excitations carry local charge 1/3;  its excitations obey Abelian braid statistics; it is topologically distinct from the usual 1/3 state; it has a complex edge with multiple channels; and its edge does not support, in the absence of reconstruction, backward moving neutral modes.  

While the WYQ states clearly represent a new kind of order, one may ask if they are at all realizable. The interaction $V_m=\delta_{m,3}$ appears unphysical, because it penalizes pairs with relative angular momentum $m=3$ but has no repulsion in the angular momentum $m=1$ channel. However, precisely this interaction is realized for composite fermions in the second $\Lambda$L! Refs. \cite{LeeScarola,Sitko,Lee1} have shown that for composite fermions in this filling factor region, the $V_3$ pseudopotential dominates (which was the motivation for WYQ's considering this interaction). To test if this physics actually underlies the 4/11 and 5/13 FQHE, we obtain the unconventional WYQ ground states $\Phi_{4/3}^{\rm uncon}$ and $\Phi_{5/3}^{\rm uncon}$ at 4/3 and 5/3 by an exact numerical diagonalization of the WYQ interaction $V_m=\delta_{m,3}$, and then composite-fermionize them to obtain explicit trial wave functions for the CF states at 4/11 and 5/13, denoted $\Psi_{4/11}^{\rm uncon}$ and $\Psi_{5/13}^{\rm uncon}$. (See SM for details \cite{SM}.) A direct comparison with the CFD ground states, shown in Table \ref{overlap}, provides strong support that these wave functions capture the physics of the actual 4/11 and 5/13 FQHE states. In other words, incompressibility at these fractions results because the occupation of CF pairs with relative angular momentum $m=3$ is minimized, distinct from the usual mechanism for FQHE at $n/(2n\pm 1)$ which minimizes occupation of {\em electron} pairs with $m=1$. 

The high overlaps of the 4/11 and 5/13 ground states with the composite-fermionized WYQ wave functions demonstrates that the 3-body terms in the inter-CF interaction do not significantly affect the nature of the 4/11 and 5/13 {\em ground} states. This is somewhat surprising because the 3-body terms are responsible for substantial differences between the excitation spectra of the corresponding systems (paired horizontally in Fig.~\ref{411}) at 4/11 and 5/13.

\begin{table}[b]
\begin{tabular}{| c | c | c | c |} \hline
$N$ & $2Q$ & $N^*$ &  $\langle \Psi_{\frac{4}{11}}^{\rm CFD}| \Psi_{\frac{4}{11}}^{\rm uncon}\rangle$     \\ \hline
12 & 28 & 5 &  1.000 \\
16 & 39 & 6  &  0.9985(1)  \\
20 & 50 & 7  &  0.9834(1)  \\
24 & 61 & 8  &  0.9351(2)  \\
28 & 72 & 9  &  0.9627(2)  \\ \hline
\end{tabular} 
\begin{tabular}{| c | c | c | c |} \hline
$N$ & $2Q$ & $N^*$ &  $\langle \Psi_{\frac{5}{13}}^{\rm CFD}| \Psi_{\frac{5}{13}}^{\rm uncon}\rangle$     \\ \hline
11 & 26 & 4 &  1.000 \\
16 & 39 & 6  &  0.9985(1)  \\
21 & 52 & 8  &  0.9962(1)  \\ 
26 & 65 & 10 &  0.9875(2) \\
31 & 78 & 12 &  0.9428(3)\\\hline
\end{tabular}
\caption{Testing trial wave functions for the unconventional states at 4/11 and 5/13. This table gives the overlap between $\Psi^{\rm uncon}$ and $\Psi^{\rm CFD}$ for 4/11 and 5/13. 
The trial wave functions $\Psi_{4/11}^{\rm uncon}$ and $\Psi_{5/13}^{\rm uncon}$ are derived by composite-fermionization of $\Psi_{4/3}^{\rm uncon}$ and $\Psi_{5/3}^{\rm uncon}$, which, in turn, are exact ground states of the WYQ interaction that select states that minimize the occupation of pairs with relative angular momentum 3. $\Psi_{4/11}^{\rm CFD}$ and $\Psi_{5/13}^{\rm CFD}$, obtained by CF diagonalization, are essentially exact. The spherical geometry is used. $N$ is the total number of electrons or composite fermions, $N^*$ is the number of composite fermions in the second $\Lambda$L, and $2Q$ is the number of flux quanta passing through the surface of the sphere.}\label{overlap}
\end{table}

Our work has a number of experimental implications. To begin with, it implies that fully spin polarized FQHE is possible at 4/11 and 5/13 under appropriate conditions. The analogy to the WYQ states implies that the 4/11 (5/13) state does not involve pairing, supports charge 1/11 (1/13) excitations with Abelian braid statistics, has multiple edge channels, and does not have (has) backward moving neutral modes.  The absence of a well defined magneto-roton branch in the finite-system spectra indicates that their quasiparticles and quasiholes are large and complex, as has been found even for the 7/3 state \cite{Balram13}.

We now show that the electron spin also plays an interesting role. A ``conventional" partially spin polarized 4/11 state has been proposed in the past \cite{Park00,Chang-spin,Wojs-spin}, wherein composite fermions fill lowest spin-up $\Lambda$L completely and form a conventional 1/3 state in the spin reversed {\em lowest} $\Lambda$L, giving a polarization $\gamma=(\nu^*_\uparrow-\nu^*_\downarrow)/(\nu^*_\uparrow+\nu^*_\downarrow)=1/2$, where $\nu^*_\sigma$ represents the filling factor of composite fermions with spin $\sigma$. The conventional mechanism for the partially spin polarized state has been confirmed by CFD \cite{Chang-spin}.  The interaction energy of the partially polarized ground state \cite{Park00,Chang-spin}, $-0.4205(2)\,e^2/\epsilon \ell$, is less than that of the fully spin polarized state, $-0.4141(2)\, e^2/\epsilon \ell$ (Fig. \ref{gaps}), indicating that the partially polarized state is stabilized at sufficiently low Zeeman splitting $E_{\rm Z}$, defined as the energy required to flip a single spin. Equating the per-particle Coulomb energy difference to $E_{\rm Z}/4$ (as 1/4 of the composite fermions flip their spin in going from fully to partially polarized state), a phase transition from the partially spin polarized conventional state to a fully spin polarized unconventional state is predicted to occur at $\kappa\equiv E_{\rm Z}/(e^2/\epsilon \ell)=0.025$. For GaAs parameters (band mass $m_{\rm b}=0.067 m_{\rm e}$, Land\'e g factor $g=-0.44$, background dielectric function $\epsilon=13.6$), this translates into a transition at a magnetic field $B_{\rm crit}\sim 19$T. (Finite width corrections are not considered explicitly here, but are expected to reduce $B_{\rm crit}$ by 10-20\% \cite{Archer13}.) Our detailed calculations thus lead to the surprising prediction, at variance with the earlier conclusion \cite{Pan03}, 
that the 4/11 state observed in Ref.~\cite{Pan03} is partially spin polarized with  $\gamma=1/2$ even though it occurs at a magnetic field as high as $\sim 11$T.  (The insensitivity of resistance to variations in $E_{\rm Z}$ \cite{Pan03} can be explained by noting that the lowest gap in the partially polarized state corresponds to an excitation within the spin reversed sector \cite{Park00,Chang-spin}, and therefore does not involve a spin reversal.) An experimental verification of this predictions, as well as of a magnetic transition at $\kappa\approx 0.025$ (for $\nu=4/11$), will serve as nontrivial confirmations of the physics described above.  The spin polarizations and spin phase transitions at $\nu=n/(2pn\pm 1)$ have been measured by transport \cite{Eisenstein,Duspin,Yeh,Cho,Eom, Betthausen}, optical \cite{Kukushkin,Yusa01,Gros07,Hayakawa}, NMR \cite{Melinte,Smet01,Kraus,Smet,Tiemann}, and compressibility \cite{Yacoby2} measurements; analogous valley polarization transitions have been observed in AlAs quantum wells \cite{Shayegan1,Shayegan2,Shayegan3}; and the experimental observations are generally consistent with the CF theory \cite{Park98, Archer13}. 

We have also evaluated the pair correlation functions as well as the density profiles of the quasiparticle and quasihole for the conventional and the unconventional states. The differences between them for conventional and unconventional states are substantial for 1/3 but less so for 4/11, as shown in the SM \cite{SM}.

It is interesting to ask what other analogous unconventional CF liquids are possible. We have investigated this question by diagonalizing both the second $\Lambda$L interaction given in Ref.~\cite{LeeScarola} and the model $V_m=\delta_{m,3}$ interaction for a wide range of particle numbers and flux values, and found that, in the range $2>\nu^*>1$, it produces incompressibility only at $\nu^*=4/3$, 5/3, 6/5 and 9/5. To the extent this model is applicable, our study implies that unconventional CF states occur at 4/11, 5/13, 6/17 and 9/23, which, along with 3/8 \cite{Pan03,MukherjeeMandalWojs2012}, exhaust all possible FQHE in the range $2/5 > \nu >1/3$ for a fully spin polarized system. 

We have thus shown that the fully spin polarized FQHE at 4/11 and 5/13 originates from a novel mechanism, due to a peculiar interaction between composite fermions. We have predicted that the previously observed state at 4/11 \cite{Pan03} is partially spin polarized, and that a transition into a fully polarized state will occur at $\kappa\equiv E_{\rm Z}/(e^2/\epsilon \ell)\approx 0.025$.  We close with a further remarkable implication of our study. It is well appreciated that the nature of the FQHE depends sensitively on the interaction pseudopotentials. That is the reason why FQHE in the second LL of GaAs is different from that in the lowest LL, and no FQHE is seen in yet higher LLs; that is also partly responsible for differences between FQHE in GaAs and graphene, and between FQHE of electrons and hard core bosons. Composite fermions in various $\Lambda$Ls provide yet another system of particles with rather unusual interactions \cite{Sitko,Lee1,LeeScarola}, which can possibly spawn new unconventional quantum liquids. The higher $\Lambda$Ls of composite fermions are likely to serve as a playground for the possible discovery of new topological states as the sample quality improves in the coming years.

We are grateful to A. C. Balram, T. H. Hansson, J. J. Quinn and S. H. Simon for fruitful discussions. We acknowledge financial support from CSIR, Government of India (SM); the DOE grant no. DE-SC0005042 (YHW,JKJ); and the Polish NCN grant 2011/01/B/ST3/04504 and EU Marie Curie Grant PCIG09-GA-2011-294186 (AW). The computations were performed using computing facilities of the Department of Theoretical Physics at IACS, Cyfronet and WCSS (both parts of PL-Grid Infrastructure), and the Research Computing and Cyberinfrastructure, a unit of Information Technology Services at Pennsylvania State University. YHW thanks the authors of the DiagHam package for sharing their programs.

\end{document}




\setcounter{figure}{0}
\setcounter{equation}{0}
\renewcommand\thefigure{S\arabic{figure}}
\renewcommand\thetable{S\arabic{table}}
\renewcommand\theequation{S\arabic{equation}}
\renewcommand\thepage{S\arabic{page}}

\begin{center}
 {\large \bf Supplemental Material for ``Enigmatic 4/11 State: A Prototype for Unconventional Fractional Quantum Hall Effect"}
\end{center}

\section{Relation between shifts}

In the spherical geometry \cite{Haldane1}, we consider $N$ electrons moving on the surface of the sphere, with a radial magnetic field that produces a total flux of $2Q$, in units of $\phi_0=hc/e$. This maps into a system of $N$ composite fermions at $2Q^*=2Q-2(N-1)$, assuming two vortices bound to each composite fermion. We will be interested in situations where composite fermions fill the lowest $\Lambda$L completely and the second $\Lambda$L partially.  We define $N^*_{\phi}=2(Q^*+1)$, so that the degeneracy of the second $\Lambda$L is $N^*_\phi+1$. The number of composite fermions in the second $\Lambda$L is 
\begin{eqnarray} 
N^*=N-(2Q^*+1)
\end{eqnarray}
Let us consider the state at $\nu^*$ with shift $S^*$, which occurs at 
\begin{eqnarray}
N^*_\phi=2(Q^*+1)=(\nu^*)^{-1}N^*-S^*
\end{eqnarray}
Eliminating $N^*$ gives 
\begin{eqnarray}
2Q^*=(1+\nu^*)^{-1}(N-1)-(S^*+2)\nu^*(1+\nu^*)^{-1}
\end{eqnarray}
which gives the electronic state at 
\begin{eqnarray}
2Q&=&\nu^{-1}N-S\\
\nu&=&{1+\nu^*\over 3+2\nu^*}\\
S &=& {3+\nu^*(S^*+4)\over 1+\nu^*} \label{SS*}
\end{eqnarray}
This equation relates the shifts at 1/3 and 2/3 (denoted $S^*$) to the shifts at 4/11 and 5/13. 

\section{CF diagonalization (CFD)}

The method of CF diagonalization~\cite{Mandal1} obtains CFD spectra and CFD eigenfunctions by a diagonalization of the full Coulomb interaction within a correlated CF basis, which is much smaller than the full Coulomb basis (some dimensions given below). The method has been described in detail elsewhere, but we give here an outline for completeness. It proceeds along the following steps:

(i) We first construct the basis \{$\Phi^{\alpha}_{2Q^*,L}$\} of all degenerate states of noninteracting fermions at $2Q^*$ in the lowest kinetic energy band with total orbital angular momentum $L$; $\alpha$ labels different states at a given $L$. For $\nu^*=4/3$ and 5/3, these states have the lowest LL fully occupied and the second LL partially occupied; the basis \{$\Phi^{\alpha}_{2Q^*,L}$\} contains all possible arrangements of fermions in the second LL. Eigenstates of definite total angular momentum $L$ can be constructed by group theory for small $N^*$, but, in general, we find it convenient to diagonalize any 2-body interaction to obtain basis functions with definite $L$.

(ii) Next we composite-fermionize this basis to obtain a basis \{$\Psi^{\alpha}_{2Q,L}$\} at $2Q$ defined as
\begin{eqnarray}
\Psi^{{\rm CF},\alpha}_{2Q,L}={\cal P}_{\rm LLL}J^2 \Phi^{\alpha}_{2Q^*,L}
\end{eqnarray}
where
\begin{eqnarray}
J^2=\prod_{j<k}(u_iv_j-v_iu_j)^2
\end{eqnarray}
is the standard Jastrow factor that attaches two vortices to each electron to convert it into a composite fermion, and ${\cal P}_{\rm LLL}$ represents projection into the LLL. The spinor coordinates are defined as
\begin{eqnarray}
u_j= \cos (\theta_j/2)\exp (-i\phi_j/2), \; v_j=\sin (\theta_j/2) \exp (i\phi_j/2)
\end{eqnarray}
where $\theta$ and $\phi$ are the angular coordinates on the surface of the sphere. The LLL projection is carried out by the method of Ref.~\cite{JainKamilla1}.

(iii) We finally orthogonalize the CF basis and diagonalize the full Coulomb interaction in this basis. The orthogonalization and the evaluation of the Coulomb matrix elements are accomplished by the Monte Carlo method of Ref.~\cite{Mandal1}. This produces the CFD spectrum as well as the CFD eigenfunctions.

\section{Trial wave functions for the conventional and unconventional states}

We construct states $\Phi_{4/3}^{\rm uncon}$ and $\Phi_{5/3}^{\rm uncon}$, wherein the LLL is fully occupied and the electrons in the second LL form the unconventional  1/3 and 2/3 states. We then composite fermionize these states to obtain explicit trial wave functions for the unconventional CF states at 4/11 and 5/13: 
\begin{eqnarray}
\Psi_{4/11}^{\rm uncon}={\cal P}_{\rm LLL} J^2 \Phi_{4/3}^{\rm uncon}
\label{eq411}
\end{eqnarray}
\begin{eqnarray}
\Psi_{5/13}^{\rm uncon}={\cal P}_{\rm LLL} J^2 \Phi_{5/3}^{\rm uncon}
\end{eqnarray}
The conventional states at 4/11 and 5/13 are constructed analogously. 
The shifts of various conventional and unconventional states, obtained with the help of Eq.~\ref{SS*} (with the shifts at 1/3 and 2/3 denoted by $S^*$), are given in the main text. 

An important quantity for a liquid is its pair correlation function $g(\vec{r})$, namely  
the probability of finding two particles at a distance $\vec{r}$ in the ground state of the system. It is defined as
\begin{eqnarray}
g(\vec{r}) = \frac{1}{\rho N}  \langle\sum_{i\neq j} \delta^{(2)}(\vec{r}-\vec{r}_{ij})\rangle \,\, ,
\end{eqnarray}
where $\vec{r}_{ij} = \vec{r}_i-\vec{r}_j$ is taken as the arc distance between two particles at the positions $\vec{r}_i$ and $\vec{r}_j$,
$\rho$ is the average density of electrons, and  the angular brackets represent the expectation value in the respective ground states.

The quasiparticle and quasihole of the conventional 1/3 state occur at $2Q=3N-4$ and $2Q=3N-2$, and have total angular momentum $L=N/2$. We find numerically that, for the WYQ interaction $V_m=\delta_{m,3}$, the lowest energy states at $2Q=3N-8$ and $2Q=3N-6$ also occur at $L=N/2$, which we identify as its quasiparticle and quasihole of the unconventional 1/3 state.  The quasiparticle and quasihole localized at the North Pole can be obtained by diagonalizing $V_m=\delta_{m,3}$ in the sector $L_z=L=N/2$, where $L_z$ is the z-component of the total angular momentum. The wave functions for the localized quasiparticle and quasihole of the 4/11 state are constructed from these by composite-fermionization as in Eq.~\ref{eq411}. 

\begin{center}
\begin{figure}[ht]
\centering
\includegraphics[scale=0.4,angle=270]{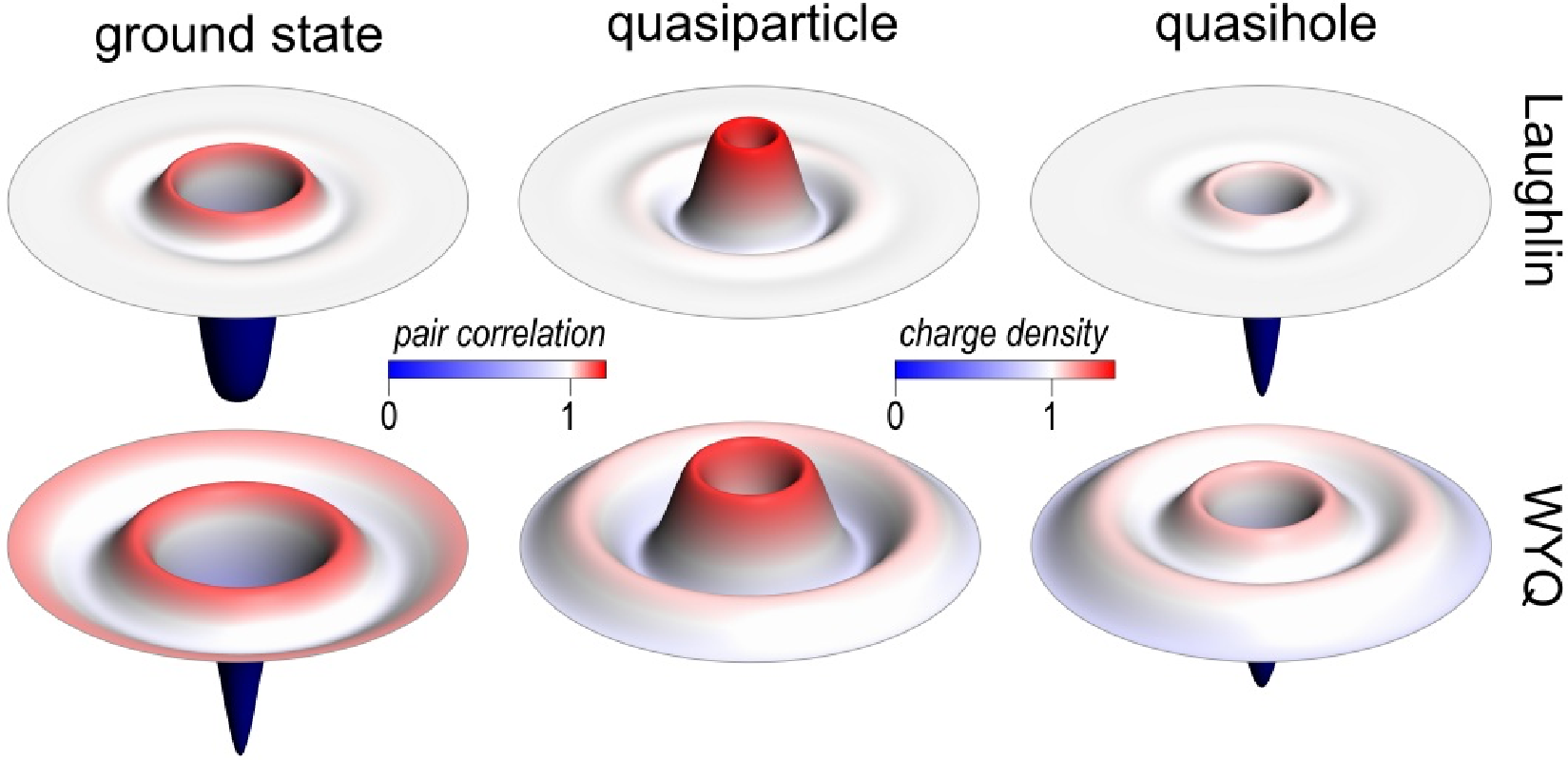}
\caption{Pair correlation function and density profiles of the quasiparticle and quasihole at 1/3. The left panels show the pair correlation function for the ground state at 1/3. The middle panels show the density of a quasiparticle located at the origin. The right panels show the density of a quasihole located at the origin. The upper row is for the conventional state, and the lower row for the unconventional state.  The results are for a system with 15 particles, and the radius of each disk is 15 magnetic lengths. All quantities are normalized so that they approach unity at large $r$. The quasiparticle (quasihole) wave functions at 1/3 are defined as the the states obtained by removal (addition) of one flux quantum, and picking out the state with $L=N/2=L_z$. The distances are quoted in units of the magnetic length $\ell$.} \label{WYQL13}
\end{figure}
\end{center}

\begin{figure}
\includegraphics[scale=0.4,angle=270]{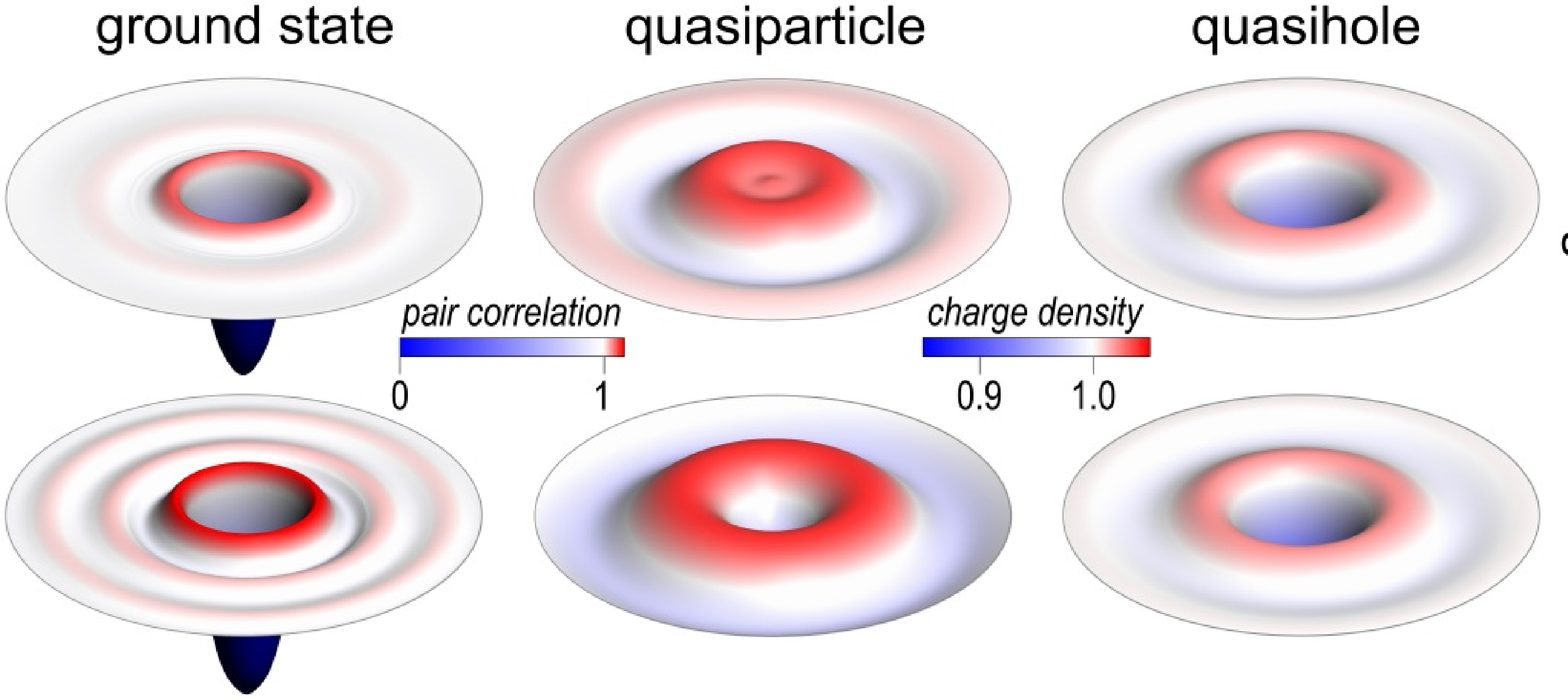}
\caption{Pair correlation function and density profiles of the quasiparticle and quasihole at 4/11. The left panels show the pair correlation function for the ground state at 4/11, defined as the probability of finding two electrons at a distance $r$ (measured in units of the magnetic length $\ell$). The middle panels show the density profile of a quasiparticle located at the origin. The right panels show the density of a quasihole located at the origin. The upper row corresponds to the conventional 4/11 state, and the lower to the unconventional 4/11 state. The results are for a system with 32, 31, and 33 particles respectively, and the radius of each disk is 15 $\ell$. All quantities are normalized so that they approach unity at large $r$.} \label{all}
\end{figure}

Fig.~\ref{WYQL13} shows the pair correlation functions for the conventional and unconventional states at 1/3, as well as the density profiles of the quasihole and quasiparticle at the origin. The analogous quantities at 4/11 are shown in Fig.~\ref{all}. 

\section{Hilbert space dimensions}

An advantage of the CFD is that the dimension of the basis \{$\Psi^{{\rm CF},\alpha}_{2Q,L}$\} is exponentially small compared to that of the full LLL basis, which allows us to go to much larger systems than possible by exact diagonalization. Tables~\ref{Exact} and~\ref{Exact2} give dimensions of the full LLL Hilbert space for various systems considered here. The dimensions quoted in Table~\ref{Exact2} are estimates for the $L=0$ sector obtained by a second order polynomial fit of log of the dimension versus $1/Q$ for a given $N$.  The dimensions of the corresponding CF basis are given in Table~\ref{CFD}. These show how an exponential reduction is achieved by working with composite fermions (without compromising on accuracy, as shown in Fig. 1 of the main text).

\begin{center}
\begin{table}[ht!]
\caption{Total number of states in the lowest Landau level at various values of $N$ and $2Q$ in the sector $L=L_z=0$.}
\label{Exact}
\setlength{\tabcolsep}{0.15em}
\centering
\footnotesize
\vspace{2ex}
\begin{tabular}{|c||rrrrrrrrrrr|}
\hline
$(N,2Q)$&$L=0$&1&2&3&4&5&6&7&8&9&10\,\\ 
\hline
\hline
$(11,26)$&320&821&1452&1940&2549&3008&3580&3999&4524&4890&5355\,\\
$(12,28)$&902&2405&4182&5662&7384&8787&10437&11741&13263&14452&15835\,\\
$(16,39)$&250256&742459&1240794&1728894&2220978&2700716&3182525&3649966
         &4117437&4568814&5018251\,\\
$(20,50)$&105917976&317504512&528896826&739596937&949661316&1158594575
         &1366455834&1572753851&1777551649&1980363901&2181258782\,\\
$(21,52)$&\,364450143&1092798053&1820337393&2545967795&3269434834
         &3989644615&4706352540&5418473549&6125774526&6827187570
         &7522492598\,\\
\hline
\end{tabular}
\end{table}
\end{center}

\begin{center}
\begin{table}[ht!]
\caption{Estimate of the total number of states in the $L=0$ sector for several values of $(N,2Q)$.}
\label{Exact2}
\centering
\footnotesize
\vspace{2ex}
\begin{tabular}{|c||c|}
\hline
$(N,2Q)$&$L=0$\\ 
\hline
\hline
(24,61)&\,$4\times10^{10}$\\
(26,65)&\,$4\times10^{11}$\\
(28,72)&\,$2\times10^{13}$\\
(31,78)&\,$7\times10^{14}$\\
\hline
\end{tabular}
\end{table}
\end{center}

\begin{center}
\begin{table}[ht!]
\caption{Number of states in the basis used in CF diagonalization. This is also equal to the dimension of the Hilbert space in the lowest band for $N$ particles at $2Q^*$.}
\label{CFD}
\centering
\footnotesize
\vspace{2ex}
\begin{tabular}{|c||rrrrrrrrrrr|}
\hline
$(N,2Q)$&$L=0$&1&2&3&4&5&6&7&8&9&10\,\\
\hline
\hline
$(11,26)$ \& $(12,28)$&1&0&2&1&2&1&2&1&1&0&1\,\\
$(16,39)$&3&0&4&3&6&3&7&4&6&4&5\,\\
$(20,50)$ \& $(21,52)$&4&3&10&9&16&14&19&17&21&18&21\,\\
$(24,61)$ \& $(26,65)$&12&10&32&30&51&48&66&61&77&70&83\,\\
$(28,72)$ \& $(31,78)$&\,\phantom{0}28&\phantom{0}48&\phantom{0}99&122&169&184&232&242&278&287&317\,\\
\hline
\end{tabular}
\end{table}
\end{center}

\section{Exact diagonalization}

The exact energy spectra of Fig.~1 of the main text were obtained by a projected Lanczos method, carried out separately for each value of $L$. Specifically, to produce energy levels at a given $L$, the Hamiltonian was diagonalized in the subspace of $L_z=L$, and each Lanczos iteration was preceded by an additional whole (nested) Lanczos procedure for the operator $L^-L^+=L^2-L_z(L_z+1)$. Even with efficient parallel coding for the (very sparse) Hamiltonian and angular momentum matrices, such exact diagonalization is only possible up to $N=16$, for which dimensions of $L_z$-spaces at $2Q=39$ is approximately $7\times10^8$ and the numbers of nonzero hamiltonian and $L^-L^+$ matrix elements exceed $4\times10^{11}$ and $6\times10^{10}$, respectively. Our highly optimized code ran this system at a speed of approximately one Lanczos iteration per 4 hours on a 12-core dual Xeon X5650 2.67GHz computing node. Hence, producing the $(N,2Q)=(16,39)$ spectrum of Fig.~1(b,d) took us about two months of uninterrupted computing on full 11 nodes, at the computing cost of (11 values of $L$) $\times$ (200 iterations to reach convergence) $\times$ (4 hours) $\times$ (12 cores), which exceeds 100 000 compute hours or 12 compute years. Since computation of excited states requires full orthogonalization, storage of many huge Lanczos vectors is also problematic. For the spectrum in Fig.~1(b,d) we reached the total disk usage of (11 values of $L$) $\times$ (200 vectors) $\times$ (dimension $7\times10^8$) $\times$ (8 bytes per double precision value), which exceeds 11 Terabytes. The CF diagonalization for the same system took approximately 200 compute hours to obtain a reasonable accuracy. 

\section{Nature of the WYQ state}

The $V_3$ interaction gives an incompressible state at $\nu=1/3$ at shift 7, which we refer to as the WYQ state. We have not been able to construct an accurate trial wave function for this state. However, many of its properties can be ascertained from exact results, which are expected to carry over to the 4/11 state since it is a WYQ state of composite fermions. We first list these statements and then explain in detail how they are derived from the numerical results obtained using a combination of spherical, torus and disk geometries.

\begin{center}
\begin{table}[ht!]
\caption{Angular momenta of quasiparticle and quasihole excitations.}
\label{SphereExcitedState}
\centering
\footnotesize
\vspace{2ex}
\begin{tabular}{|c||rrrrrrrrrrr|}
\hline
$N$  &6&7&8&9&10&11&12&13&14&15&16\,\\
\hline
\hline
$L_{\rm qp}$ &3&3.5&6&6.5&5&5.5&6&6.5&7&7.5&8\,\\
$L_{\rm qh}$ &1&1.5&4&4.5&5&5.5&6&6.5&7&7.5&8\,\\
\hline
\end{tabular}
\end{table}
\end{center}

\begin{center}
\begin{figure}[ht]
\centering
\includegraphics[scale=0.6]{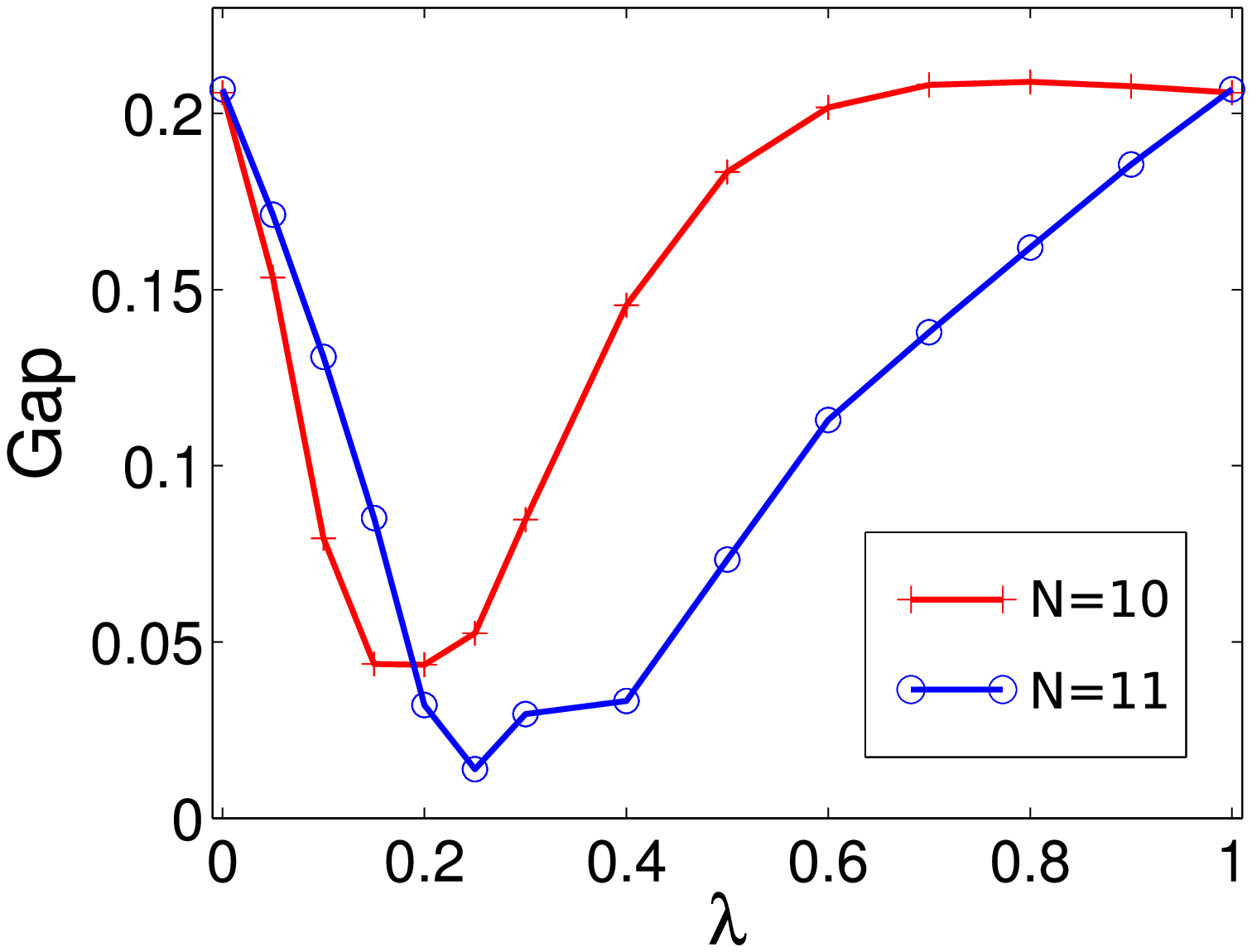}
\caption{Evolution of the gap in energy spectrum when tuning the parameter $\lambda$ in the Hamiltonian Eq.~\ref{V1V3Ham}} 
\label{GapValue}
\end{figure}
\end{center}

\begin{center}
\begin{figure}[ht]
\centering
\includegraphics[width=0.85\textwidth]{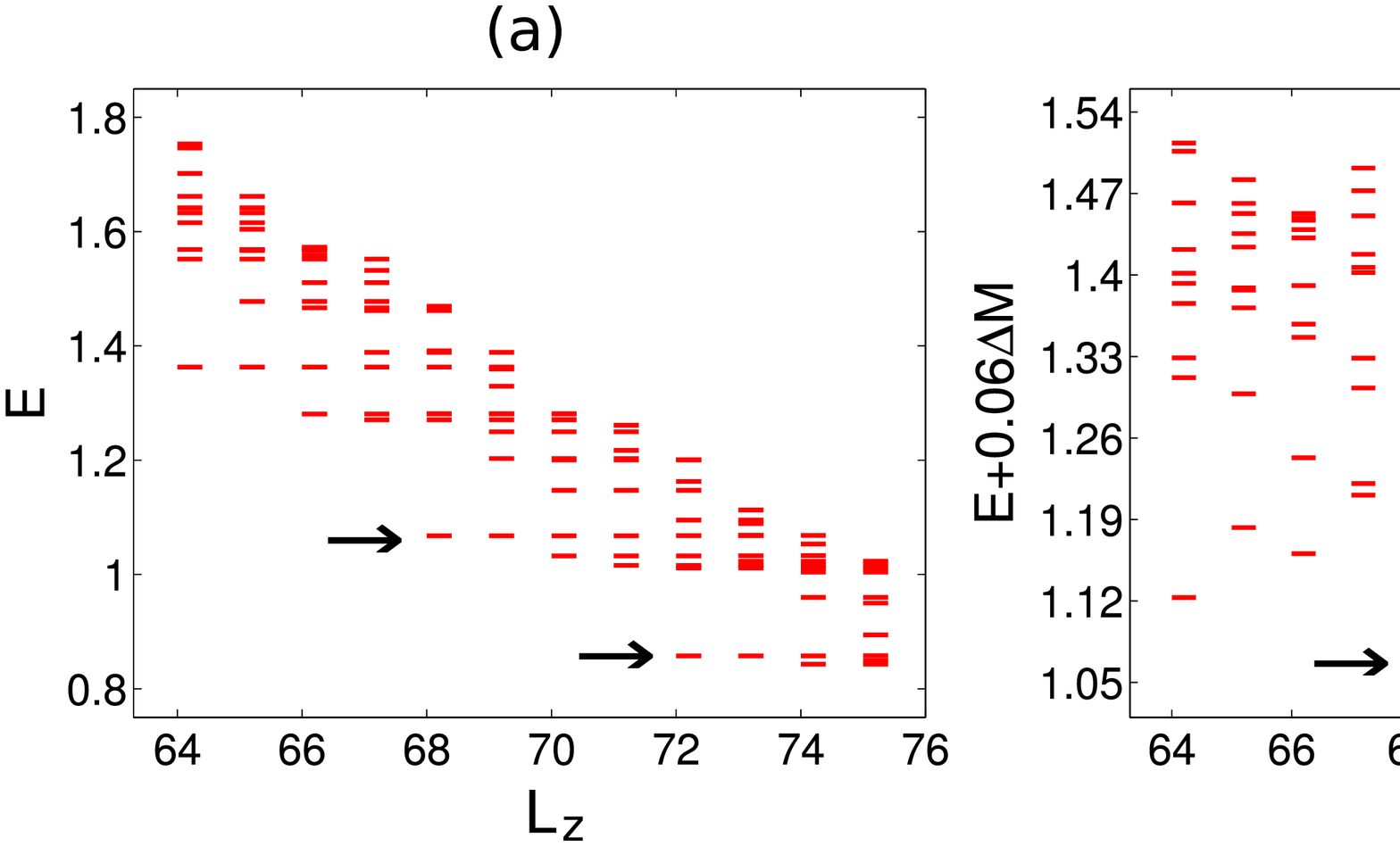}
\caption{Energy spectrum on disk which shows the edge excitations of the WYQ state with $N=8$ and ${\Delta}M=L_z-68$.} 
\label{DiskEnergy}
\end{figure}
\end{center}

\begin{center}
\begin{figure}[ht]
\centering
\includegraphics[width=0.85\textwidth]{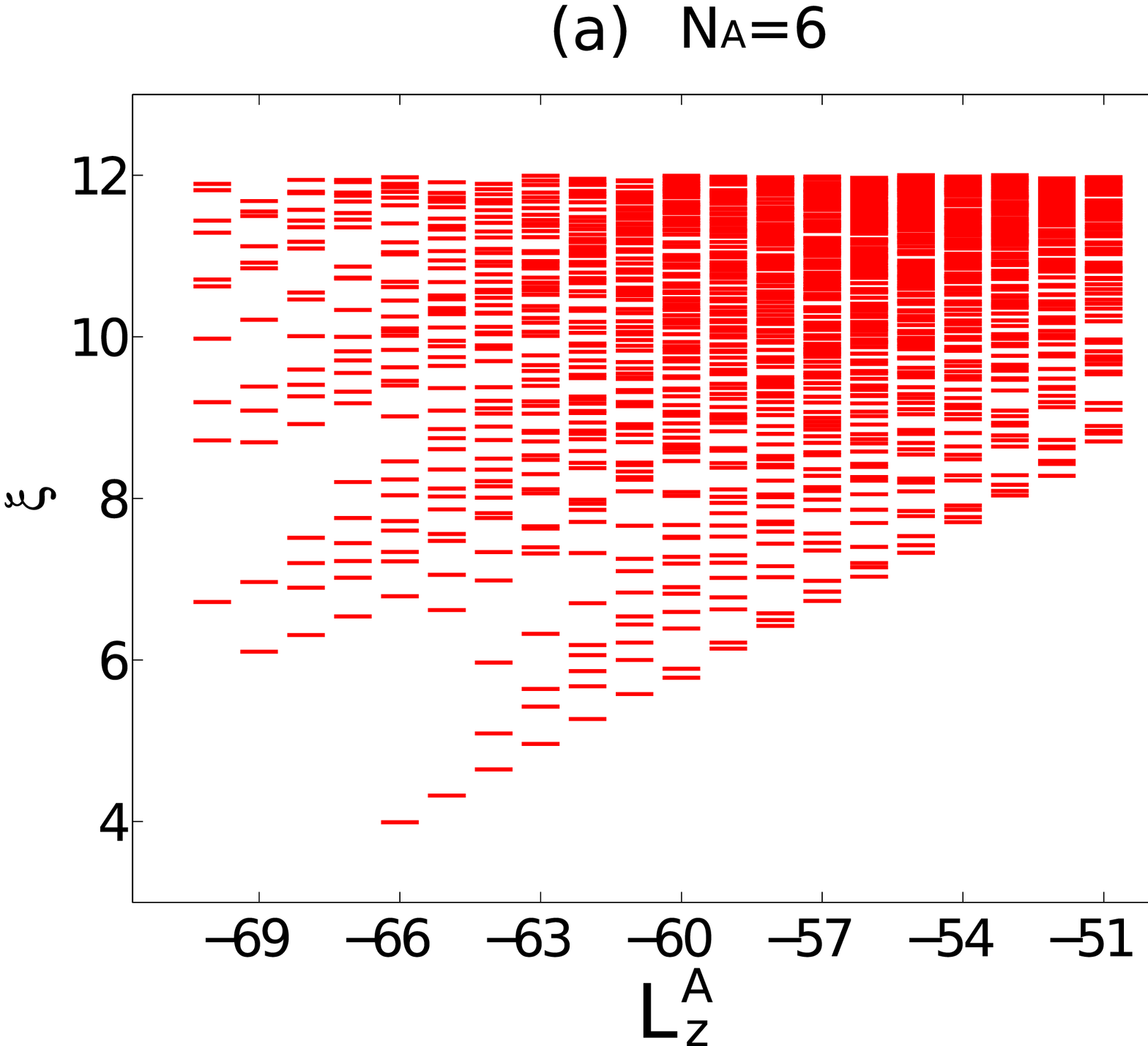}
\caption{Real space entanglement spectrum of the WYQ state on sphere with $N=14$ and $2Q=46$.} 
\label{RealSpaceEntangle}
\end{figure}
\end{center}

(i) The WYQ state does not involve pairing because an incompressible state occurs for both even and odd numbers of particles.

(ii) The excitations of the WYQ state have local charge $1/3$ in units of the electron charge. This can be seen by analyzing the energy spectra at $2Q=3N-7$, {\em i.e.} for $2Q=3N-6$ (quasihole) and $3N-8$ (quasiparticle). As shown in Table~\ref{SphereExcitedState}, these states occur at total angular momentum $L=N/2$ for $N{\geq}10$, which is analogous to the behavior for the ordinary 1/3 state. This implies that a single quasiparticle or quasihole is created when the flux is changed by one unit, which thus has a charge of magnitude $\nu=1/3$. (If addition or subtraction of one flux quantum created more than one quasihole or quasiparticle, a band consisting of several quasi-degenerate states would emerge.) 

(iii) The excitations of the WYQ state obey Abelian braid statistics. This claim is supported by the fact that the WYQ state has no pairing as well as our calculations on torus that show precisely three degenerate ground states (see below).

(iv) The WYQ state is topologically distinct from Laughlin state. One clear evidence in support of this assertion is that these two states occur at different shifts on sphere. The topological distinction between these two states also implies that the transition between them should be accompanied by a gap closing. This is most conveniently investigated in the torus geometry ~\cite{Yoshioka}, because the fact  that they occur at different shifts on sphere prevents a direct study of the phase transition between them. To this end, we consider a rectangular torus spanned by lattice vectors ${\mathbf L}_1=L_1{\widehat e}_x$ and ${\mathbf L}_2=L_2{\widehat e}_y$ with $L_1=L_2$. We use the Landau gauge ${\mathbf A}({\mathbf r})=Bx{\widehat e}_y$ to generate a magnetic field ${\mathbf B}=B{\widehat{\mathbf z}}$. The magnetic translation operators are defined as $T({\mathbf d})=e^{-i{\mathbf d}\cdot{\mathbf K}}$ with ${\mathbf K}=-i\hbar\nabla-e{\mathbf A}+e{\mathbf B}\times{\mathbf r}$. The periodic boundary conditions $T({\mathbf L}_\alpha)=1$ quantize the flux $N_\phi=L_1L_2/(2{\pi}\ell^2_B)$ to be an integer multiple of flux quanta with $\ell_B=\sqrt{\hbar/eB}$. The usual basis in the lowest Landau level is 
\begin{eqnarray}
\psi_\alpha(x,y) = \frac{1}{(\sqrt{\pi} L_2 \ell_B)^{1/2}} \sum_n^{\mathbb{Z}} \exp\Big[ 2\pi(\alpha+nN_\phi)\frac{x+iy}{L_2} - \frac{\pi L_1} {N_\phi L_2} (\alpha+nN_\phi)^2 \Big] e^{-x^2/(2\ell^2_B)}
\end{eqnarray}
where the index $\alpha$ gives the momentum $k_y=2{\pi}\alpha/L_2$ along the $y$ direction. For an eigenstate of translationally invariant Hamiltonian, $Y=\sum^{N}_{i=1} \alpha_i$ mod $N_\phi$ is a conserved quantity that can be used to label the state. A more complete symmetry analysis shows that two many-body translation operators~\cite{Haldane2} can be defined and the quantum numbers $\kappa_x$ and $\kappa_y$ associated with these operators provide more information about the nature of a state. In particular, a rotationally invariant incompressible state should occur at $\kappa_x=\kappa_y=0$ in finite size systems. As in the spherical geometry, many-body interaction can be parametrized using pseudopotentials which we denote as $V_m$. When the interaction is chosen such that only $V_1$ is non-zero, the Laughlin state is produced as the exact eigenstate with a 3-fold ground state degeneracy. We obtain a WYQ ground state with 3-fold degeneracy when only $V_3$ is non-zero. Varying $\lambda$ from $0$ to $1$ in the Hamiltonian 
\begin{eqnarray}
H = (1-\lambda) V_1 + \lambda C V_3
\label{V1V3Ham}
\end{eqnarray}
induces a transition from the Laughlin state to the WYQ state. We have introduced a constant $C$ to make sure that the states at the two ends, $\lambda=0$ and $\lambda=1$, have the same gap. The evolution of gap with respect to $\lambda$ is shown in Fig.~\ref{GapValue}. Note that we are defining the gap as the energy difference between the first excited state and the ground state. The gap clearly becomes small at some point, consistent with the expectation that it would vanish at some $\lambda$ in the thermodynamic limit.

(v) The edge spectrum of the WYQ state appears to contain multiple forward moving branches; {\em each} has a state counting of 1, 1, 2, 3, ... and thus constitute a Luttinger liquid described by a single bosonic field. This conclusion is arrived at by a study of the edge spectrum in the disk geometry. The single-particle eigenstates are given by
\begin{eqnarray}
\psi_m(z) = \frac{1}{\sqrt{2{\pi}2^mm!}} z^m e^{-\frac{1}{4}|z|^2}
\end{eqnarray}
where $m$ gives the $z$ component of the angular momentum and $z=x+iy$ is the complex coordinate. For an eigenstate of rotationally invariant Hamiltonian, $L_z=\sum^{N}_{i=1} m_i$ is a conserved quantity that can be used to label the state. As in the spherical and torus geometries, the many-body interaction can also be parameterized by the pseudopotentials $V_m$'s. The energy spectra for $N=8$ is shown in Fig.~\ref{DiskEnergy}: in the left panel, no confinement potential is used; in the right panel, we add a parabolic confinement potential $U=0.06(L_z-68)$ to select the state at $L_z=68$ to be the global ground state. (This choice of ground state angular momentum is dictated by the sphere-disk correspondence.) The black arrows in Fig.~\ref{DiskEnergy} indicate two branches starting at $L_z=68$ and 72, each with a state counting of $1,1,2,3 \cdots$. 

We have also studied the entanglement spectrum~\cite{Li} on sphere, from which, in principle, information about the intrinsic edge excitation spectrum (without edge reconstruction) can be extracted. To calculate the entanglement spectrum, we take an incompressible WYQ ground state $|\Psi\rangle$ and partition the system into two subsystems denoted as $A$ and $B$. Defining two sets of basis states for $A$ and $B$ as $\Phi^A_\alpha$ and $\Phi^B_\beta$, the state $|\Psi\rangle$ can be written as
\begin{eqnarray}
|\Psi\rangle = \sum_{\alpha\beta} C_{\alpha\beta} |\Phi^A_\alpha\rangle \otimes |\Phi^B_\beta\rangle = \sum_{i} e^{-\xi_i/2} |\Psi^A_i\rangle \otimes |\Psi^B_i\rangle
\end{eqnarray}
where the last step is a singular value decomposition of the coefficient matrix $C_{\alpha\beta}$, and $|\Psi^A_i\rangle$ and $|\Psi^B_i\rangle$ comprise the transformed basis states for the subsystem $A$ and $B$. A plot of the eigenvalues $\xi_i$ versus the conserved quantum numbers produces the entanglement spectrum. Based on previous experience, a cut in real space~\cite{Sterdyniak,Dubail,Rodriguez1} is the most useful for revealing edge excitations in finite system. We therefore choose the subsystem $A$ ($B$) to be the southern (northern) hemisphere. For this choice of bipartition, the number of particles $N_A$ in subsystem $A$ and the total angular momentum $L^A_z$ of subsystem $A$ are conserved quantities. We show in Fig.~\ref{RealSpaceEntangle} the real space entanglement spectra for the WYQ state with $N=14$ and $2Q=46$, from which a simple edge structure can not be identified unambiguously. One reason of this ambiguity, as one has learned from the entanglement spectrum of many composite fermion states~\cite{Rodriguez1,Rodriguez2}, is the existence of several branches in the edge spectrum. Larger systems will be needed to bring out the edge physics clearly.

(vi) The spectra in Fig.~\ref{DiskEnergy} also show that the WYQ 1/3 state (and by implication the 4/11 state) does not have any backward moving edge modes (which would have produced a dispersion with a negative slope). A consequence is that its particle hole conjugate state at 2/3 (and hence the composite fermionized state at 5/13) has backward moving neutral modes.